# Giant anisotropic magnetoresistance with dual-four-fold symmetry in $CaMnO_3$/$CaIrO_3$ heterostructures

Suman Sardar[†1], Megha Vagadia[†1], Tejas Tank[1], Sarmistha Das[2], Brandon Gunn[2], Parul Pandey[3], R. Hübner[3], Fanny Rodolakis[4], Gilberto Fabbris[4], Yongseong Choi[4], Daniel Haskel[4], Alex Frano[2], and D.S. Rana[*1]

[1]*Department of Physics, Indian Institute of Science Education and Research Bhopal, M.P. 462066, India*

[2]*Department of Physics, University of California, San Diego, CA 92093, USA*

[3]*Institute of Ion Beam Physics and Materials Research, Helmholtz-Zentrum Dresden-Rossendorf, 01328 Dresden, Germany*

[4]*Advanced Photon Source, Argonne National Laboratory, Argonne, Illinois 60439, USA*

## Abstract

The realization of four-fold anisotropic magnetoresistance (AMR) in novel 3d-5d heterostructures has boosted major efforts in antiferromagnetic spintronics. However, despite the potential of incorporating strong spin-orbit coupling, only small AMR signals have been detected thus far, prompting a search for new mechanisms to enhance the signal. In this study on $CaMnO_3$/$CaIrO_3$ heterostructures, we report a unique *dual-four-fold* symmetric 70% AMR; a signal two orders of magnitude larger than previously observed in similar systems. We find that one order is enhanced by tuning a large biaxial anisotropy through octahedral tilts of similar sense in the constituent layers, while the second order is triggered by a spin-flop transition in a nearly Mott-type phase. Dynamics between these two phenomena as evidenced by the step-like AMR and a superimposed biaxial-anisotropy-induced AMR capture a subtle interplay of pseudospin coupling with the lattice and external magnetic field. Our study shows that a combination of charge-transfer, interlayer coupling, and a spin-flop transition can yield a giant AMR relevant for sensing and antiferromagnetic memory applications.

[†]*These authors contributed equally.*

[*]*E-mail:* [dsrana@iiserb.ac.in](dsrana@iiserb.ac.in)

The relativistic spin-orbit coupling (SOC) in dimensionality-controlled quantum materials with 5d metal oxides is an emergent area of research. Apart from two-dimensional Van der Waals systems, epitaxially engineered superlattices (SLs) based on complex oxides enable terminations of different symmetries at atomically sharp heterointerfaces, creating opportunities to derive unique electronic and magnetic functionalities [1-2]. In a recent development, epitaxial SLs of SOC 5d iridium oxides and correlations-governed 3d oxides exhibit spintronics functionalities based on emergent magnetic and topological properties [3-9]. Explorations of this rich phenomenology in 3d-5d SLs are underway to be best utilized in antiferromagnetic (AFM) spintronics. This offers a dissipation-minimized and ultrafast alternative to its ferromagnetic (FM) counterpart, in which spin torque and anisotropic magnetoresistance (AMR), useful for data writing and reading operations, respectively, emerge from entangled spin-orbit interactions [10-15].

The 5d oxides $SrIrO_3$ (SIO) and $CaIrO_3$ (CIO) are fast emerging focus points for AFM spintronics platforms [3-9]. Their ground state, comprising the Dirac semimetal phase and a non-magnetic ground state of Ir pseudospin moments ($J_{eff} = 1/2$), is uniquely defined by an interplay of SOC and correlations (U) [16-19]. Tilting the balance of these energetics alters their ground state. For instance, lowering the dimensionality increases the "U" and tends to induce pseudospin-based emergent magnetism. Using this template, CIO/SIO sequences have been integrated in heterostructures to unravel magnetic phenomena potentially useful for a four-fold AMR functionality, derived from either AFM/AFM or AFM/FM-type SLs [4-5]. For example, the large-bandwidth SIO interfaced with magnetic or non-magnetic 3d systems yields two- and four-fold-AMR, owing to a rich variety of emergent magnetic phases, interface coupling, charge-transfer across the interface, and Rashba SOC [4-5, 8]. In the case of SLs of low-bandwidth CIO and $SrTiO_3$ (STO), the mechanism underlying the AMR is a combination of in-plane biaxial magnetic anisotropy, magneto-elastic coupling, and interlayer exchange coupling based on tilted oxygen octahedra with glazer notation ($a^-a^-c^+$) across the constituent layers [8, 20-21]. Despite all these diverse yet concerted efforts, the maximum amplitude of the four-fold AMR signal in 3d-5d heterostructures is limited to one percent. This demands the development of novel strategies in terms of the choice of constituents and the architecture of the heterostructures that efficiently translate the emergent magnetism into a larger AMR effect.

In this study, we demonstrate that SLs coupling the CIO with the low-bandwidth, severely distorted, and canted AFM manganite $CaMnO_3$ (CMO), wherein both constituents have the same sense of oxygen octahedra tilts ($a^-a^-c^+$), elevates the AMR by more than one order of magnitude to ~10%. In addition, the spin-flop transition in a nearly Mott region emerges as a new mechanism to trigger a further increase of the AMR amplitude by almost an order of magnitude along with a second four-fold symmetric component. We show that these two effects converge to yield an unprecedented AMR of about 70%, two orders of magnitude larger than achieved to date, and discuss their control mechanisms and other unique facets.

The SLs $[(CMO)_x/(CIO)_y]_z$ (x, y = number of unit cells (UC)/period; z = repetitions) used in this study are labelled as $(MIxy)_z$, where M and I refer to the CMO and CIO layers, respectively. These SLs were placed in different categories based on their construction, period of constituent layers, and intra- and interlayer coupling designed to obtain a large AMR (details appended in the Supplementary Information). The quality of the SLs was explored by x-ray diffraction techniques and high-angle annular dark-field scanning transmission electron microscopy [HAADF-STEM] and is discussed in the Supplementary Information (Figures S1 & S2). The AMR, also referred to as angular-dependent magnetoresistance, was measured in three different senses of rotation of the SLs with respect to the magnetic field (Fig. 1) [22] and is calculated as

$$\text{AMR} = \frac{\rho[B(\alpha)] - \rho[B(\alpha = 0)]}{\rho[B(\alpha = 0)]} \quad (1)$$

where α represents the angle between the magnetic field and the current direction. The in-plane and out-of-plane axes and the current directions are rotated with respect to a fixed applied magnetic field. Figure 1c depicts three rotation angles, namely $\theta$, $\gamma$, and $\phi$. The angles $\theta$ and $\gamma$ are distinguished by the direction of the current with respect to the field, which differs by 90°. Their out-of-plane axis rotates in the same direction. For the third rotation $\phi$, the in-plane axis rotates with respect to the field.

First, we discuss the magnetic and electrical behavior of $(MIxy)_z$ (x = y = 2-4), as depicted in Fig. 1 a & b. All SLs exhibit a magnetic transition in the range of 70–100 K. The transition temperatures and the saturation magnetic moments decrease with decreasing the period of the SLs (Fig. 1 a). A saturation moment of ~0.4 $\mu_B$/f.u. for $(MI22)_{10}$ suggests a canted AFM state. For comparison, the magnetization data for (MI84)$_5$ is plotted in Fig. 1b which does not show a magnetic transition, suggesting a role played by the thickness of the manganite layer. Furthermore, while the sheet resistance increases as the CIO period decreases, the resistance of a 10 nm CIO film is clearly different from those of the SLs. The (MI22)$_{10}$, in particular, tends to exhibit a Mott-type state below 30 K. This behavior can be explained in the framework of dimensionality-induced enhancement of correlations "U" and charge transfer across the interface [6, 9]. These emergent magnetic and transport properties are related to the AMR of these SLs, as presented below.

A comparison of the $\phi$-, $\theta$-, and $\gamma$-AMR for the two key SLs (MI22)$_{10}$ and (MI33)$_5$ is presented in Fig. 1 d, e. The $\phi$-AMR for (MI22)$_{10}$ is dominantly four-fold, while the $\theta$- and $\gamma$-AMR are two-fold. (MI33)$_5$ behaves the same way except that the $\theta$- and $\gamma$-AMRs develop a subtle, yet discernible, four-fold component superimposed on a dominant two-fold AMR. Detailed AMR properties of (MI22)$_{10}$, (MI33)$_5$, (MI44)$_5$, and (MI84)$_5$ as a function of temperature, magnetic-field (B), and SL period unravel several remarkable features unique to the 3d-5d SLs (Fig. 2). (MI22)$_{10}$ exhibits an astounding $\phi$-AMR of 70% at 10 K and is the central message of this work (Fig. 2 a). This AMR reduces to 25% at 20 K and gradually ceases to manifest at 100 K. The $\phi$-AMR substantially decreases with increasing the SL period. In particular, it decreases to 3–4% for (MI33)$_5$ with a temperature dependence similar to that of (MI22)$_{10}$, followed by a further decline to 1% and 0.04% for (MI44)$_5$ and (MI84)$_5$, respectively. The $\theta$- and $\gamma$-AMRs also show a large amplitude of about 15% for (MI22)$_{10}$ and 2% for (MI33)$_5$. Usually, the $\theta$- and $\gamma$-AMRs are larger than the $\phi$-AMR [8, 23-24], but for (MI22)$_{10}$ and (MI33)$_5$ the opposite behavior points toward a different underlying mechanism.

The origin of the AMR can be ascertained by comparing its three types and analyzing their dependence on the magnetic field. None of the $\phi$-, $\theta$-, and $\gamma$-AMRs follow a quadratic dependence on B (Figure S4), which rules out Lorentz scattering as the origin [8]. In addition, the magnitude and phase for two-fold $\theta$- and $\gamma$-AMRs are exactly the same for (MI22)$_{10}$ (Fig. 1d). This discards the possibility of either spin-Hall MR or s-d scattering as the underlying mechanisms [22-25]. In addition, a weak emergent magnetism in these SLs suggests that the origin of the AMR could be domain scattering based on biaxial magnetic-anisotropy [8, 26-28]. The bi-axial magnetic anisotropy is controlled by an interlayer coupling with anisotropic exchange and Dzyaloshinskii–Moriya interactions [29-30]. The strength of this coupling depends on the dimension of the constituent layers and the interface coupling.

To shed light on the latter possibility, we determine the role of the interface and the individual CMO and CIO layer thicknesses by studying the $\phi$-AMR in (MIx2)$_z$ for x = 2, 4, 6, and 8 and (MI2y)$_z$ for y = 2, 4, and 5. In (MIx2)$_z$ for x = 4, 6, and 8, the $\phi$-AMRs are of comparable amplitude in the range of 0.3–0.6% (Figure S5), much smaller than that of (MI22)$_{10}$. Also, there is little change in the AMR as the thickness of CMO is increased to four UCs and above. Similarly, the role of the CIO layer thickness was ascertained in (MI2y)$_z$ for y = 2, 4, and 5. The $\phi$-AMR drops from 70% for (MI22)$_{10}$ to 1.6% for (MI24)$_5$ and 0.8% for (MI25)$_5$ (Figure S6). In this system, the AMR decreases more gradually with increasing *y* than the previous one. This suggests that to have a large $\phi$-AMR, both CIO and CMO layer thicknesses must be below a certain critical threshold. The influence of CMO and CIO thickness was also investigated for larger fixed periods of 4 UCs, as in (MIx4)$_5$ for x = 2, 4, and 8; and 5 UCs as in (MI5y)$_z$ for y = 2, 5, and 8. The trends observed (Figures S5 and S6) are similar to the two above-mentioned series but very different in the AMR magnitude. First, the AMR in these SLs falls in the range of 1.2–0.04%, and second, for larger-thickness SLs, such as (MI58)$_5$, the amplitude and phase of AMR converges with that of the CIO thin film (Figure S7).

The data presented above reveal that the large AMR only appears if both layers CMO and CIO are below a threshold thickness. Also, above a critical thickness of both constituent layers, the AMR of the SLs is similar to that of the CIO thin film of equivalent thickness. Therefore, this suggests that the $\phi$-AMR in these SLs originates from an in-plane biaxial anisotropy induced by the interlayer coupling, as follows. The inter- and intra-layer magnetic interaction can be written as [31-32]

$$H_{ij} = J_{ij}\,\vec{S_i}\cdot\vec{S_j}\ +\ \Gamma_{ij}\overrightarrow{S_i^z}\cdot\overrightarrow{S_j^z}\ +\ \vec{D_{ij}}\cdot[\vec{S_i}\times\vec{S_j}] \qquad (2)$$

The first term accounts for an isotropic AFM exchange, the second term describes the symmetric anisotropy responsible for in-plane AFM order, while the third term are Dzyaloshinskii–Moriya interactions that cants the in-plane AFM moments. The second term needs the interlayer coupling to create the in-plane magnetic anisotropy, which, in this case, is evident from the $\phi$-AMR [30]. Though originally proposed for layered Sr$_2$IrO$_4$ to explain the strength of interlayer coupling between SrIrO$_3$ layers separated by non-magnetic SrO layers, this is also relevant for engineered SLs where CIO or SrIrO$_3$ layers are separated by magnetic or non-magnetic layers [8, 9]. In CIO/CMO SLs, while a smaller CIO period possesses a larger "U" and a canted AFM phase, a small CMO period is conducive for stronger CIO interlayer coupling. This is evident from the close relation between emergent magnetism and AMR (Figures 1 and 2) known to emanate from interlayer-coupling-induced biaxial anisotropy, elaborated on as follows. First, $\phi$-AMR ceases to manifest above the canted AFM transition of ~70–100 K. Second, a large moment of ~0.4 $\mu_B$/*f.u.* for (MI22)$_{10}$ coincides with the large amplitude of all three types of AMR. Third, the trough and crest angles of the $\phi$-AMR shift considerably from the expected values of 45° and 90° to 52–55° and 92–94°, respectively, suggesting a large canting angle of CIO moments as also corroborated by a large magnetization (M). Last, the range of interlayer coupling is as large as 8 UCs through the CMO. In other words, the strength of the interlayer coupling decreases rapidly with increasing the CMO layer thickness, yet (MI82)$_5$ with a thick CMO layer displays a sizable AMR of 0.7%. This suggests that the interlayer coupling for 2 UCs CIO persists for a CMO period of as large as 8 UCs. Contrary to this, earlier studies on 3d-5d SLs showed a four-fold $\phi$-AMR manifested only for a maximum of 2 UCs of the 3d layer [3, 8]. The strong interlayer coupling we observe is expected to arise from the similar structural distortions of the two constituents, which facilitates charge transfer across the interface and spin-lattice coupling. Before exploring this further, it is imperative to present some unique features of the giant $\phi$-AMR in (MI22)$_{10}$ suggesting a spin-flop transition with pseudospin lattice coupling as a new mechanism for this effect.

At low temperatures, $\phi$-AMR in (MI22)$_{10}$ deviates from regular four-fold sinusoidal symmetry. As scrutinized in very close temperature intervals in the range of 10–25 K (Fig. 3), an additional four-fold pattern of AMR kinks manifests as a superimposed feature on the sinusoidal pattern while traversing from the *<100>* to the *<110>* family of crystallographic directions (depicted in Fig. 3). A well-defined four-fold $\phi$-AMR at 25 K transitions to a pattern with a multitude of kinks at 22, 20, and 18 K, followed by a symmetric and sharp four-fold single kink at 16 and 15 K. A smooth pattern appears at 14 K that further transforms to a sharp step-like amplitude of about 70% at 10 K along with a reversal in polarity of peak amplitude. This unprecedented $\phi$-AMR behavior is complex both in its pattern and amplitude. The amplitude shows two peaks (35% at 18 K and 70% at 10 K) and two dips (10% at 25 K and 11% at 14 K). The trough and the crest of the $\phi$-AMR are assigned to the difference in scattering by soft *<100>* and hard *<110>* axes, respectively, in a biaxial anisotropic system. In view of this, there appears to be an uneven scattering transition from the *<110>* family of directions at 25 K to the *<100>* directions at 14 K (Fig. 3 b). As demonstrated in Fig. 3, the *<100>* crystallographic family of directions show low resistance, and the *<110>* directions shows peak resistance. At 25 K, the uneven amplitude of the crests suggests an uneven scattering from the *<110>* family of hard directions, i.e. larger scattering from A' *[110]* and C' *[−1−10]* compared to B' *[−110]* and D' *[1−10]* (marked by a black dashed line in the bottom panel of Fig. 3b). In contrast, nearly the same magnitude of troughs suggests the uniform scattering from the *<100>* and *<010>* soft directions (marked by blue dashed lines in the bottom panel of Fig. 3b). At 14 K, however, this behavior reverses. Here, the *<100>* family of soft axes exhibits non-uniformity in the troughs (shown by red dashed lines). This reversal in non-uniformity of scattering from the *<110>* directions at 25 K to the *<100>* directions at 14 K is related to a systematic pattern of multiple and single $\phi$-AMR kinks with four-fold symmetry in the transition temperatures (Figures 3 c & d). All these facets of the $\phi$-AMR can be explained to originate from two key phenomena: (i) a strong biaxial magneto-crystalline anisotropy yielding the sinusoidal $\phi$-AMR of up to 20%, and (ii) a spin-flop transition that triggers the kink- and step-like metamagnetic AMR of up to 70% at a maximum field of 9T.

A competition between pseudospin-lattice (S-L) coupling and field-pseudospin coupling could explain the anomalous AMR in our CIO/CMO SLs. The S-L coupling in iridates is represented by [29]

$$H_{s-l} = \Gamma_1 \cos(2\alpha)(S_i^x S_j^x - S_i^y S_j^y) + \Gamma_2 \sin(2\alpha)(S_i^x S_j^y + S_i^y S_j^x) \quad - \quad (3)$$

$\Gamma_1$ and $\Gamma_2$ denote the energy scales of S-L coupling to the distortions along *<100>* and *<110>*, respectively, and $\alpha$ is the angle between staggered moments and the *<100>* direction. The competition between $xy$ and $x^2$-$y^2$ quadruple symmetries provides two solutions for H = 0; $\alpha$ = 45° for $\Gamma_1 > \Gamma_2$, and $\alpha$ = 0 for $\Gamma_1 < \Gamma_2$. The former is observed in $Sr_2IrO_4$ and an engineered $SrIrO_3/SrTiO_3$ SL of the similar magnetic structure, while the latter manifests for CIO/STO heterostructures. The $\phi$-AMR phase in our SLs lags by 45° when comparing these two cases [3, 8]. The solution to the present CIO/CMO SLs is at $\alpha$ = 0, as the minimum of AMR oscillation occurs along the *<100>* directions, similar to that in CIO/STO SLs. The difference of the phase lag in SIO- and CIO-based 3d-5d SLs is due to the different sense of octahedral rotations in their low-dimensional limits.

The S-L coupling is strong in the vicinity of a magnetic transition and weakens on lowering the temperature [29, 32]. The $\phi$-AMR, as reported in CIO/STO SLs, scales with the strength of S-L coupling, thus peaks around the transition temperature. This, however, is not the case in our CIO/CMO SLs, wherein the $\phi$-AMR peak for thin SLs manifests at temperatures much lower than the magnetic transition (both $(MI22)_{10}$ and $(MI33)_5$ display the maximum AMR at 10 K). In the thick SLs with low AMR (0.4–0.8%), such as in $(MIx2)_z$ (x = 4, 6, and 8), the peak shifts to slightly elevated temperatures in the range of 30–50 K (Figure S8). Overall, the AMR peaks well below the magnetic transition in both thick and thin SLs. This indicates a dominant role of field-pseudospin coupling in addition to the S-L coupling. Considering the effect of applied field on S-L coupling, it is known that the field-lattice coupling rotates the orthorhombic distortion as the in-plane axis is rotated with respect to the field for $\phi$-AMR measurement. At low temperatures, however, the stiffness of the lattice weakens the S-L coupling. This tilts the balance in favor of field-pseudospin coupling in the presence of large magnetic moments [29-30]. A temperature-dependent competition between these two couplings determines the $\phi$-AMR for $(MI22)_{10}$ as (i) re-orientation of moments via S-L coupling at high temperature, and (ii) direct coupling of field-spins at low temperature when the lattice is rigid but possesses larger moments.

For the SL having larger canted pseudospin moments, the transition from (i) to (ii) onsets at 25 K, below which the AMR increases significantly. This is further supported by the $\phi$-AMR with multiple kinks setting in at 22 K, which represents lower resistance and higher magnetization corresponding to the kinks rather than to the soft <100> axes. Such a state in AFM/canted AFM can be induced by a spin-flop metamagnetic transition triggered by high magnetic fields [33]. A similar spin-flop transition also occurs in the CMO/CIO SLs with the canted AFM phase as evident from well-defined metamagnetic kinks appearing in $\phi$-AMR only at the maximum field of 9 T (Figures 3e and S9). These kinks become more pronounced upon lowering the temperature from 22 to 15 K [figures 3 (c and d)] while retaining an overall sinusoidal envelope emerging from the S-L coupling (explained by eq. 3). Eventually at 10 K and in high field of 9 T, the $\phi$-AMR oscillations reverse their phase by 90°. This is highlighted by the dashed lines matching the $\phi$-AMR at (i) 10 K in fields of 5 T and 9 T (Fig. 3e), and (ii) 14 and 10 K in a field of 9 T (Figure

S10). The comparison reveals that as the component of field gets stronger along *<110>* [Fig. 3e], an avalanche-like metamagnetic transition induces a giant $\phi$-AMR. The balance shifts dominantly in favor of field-pseudospin coupling that prevents the restoration of the originally sinusoidal resistance.

Metamagnetic transitions are known to occur in manganites as well as in iridates [29, 31, 34-35]. In manganites, a transformation from a smooth to a sharp step-like metamagnetic transition manifests in both magnetization and electrical resistivity. In iridates, dimensionality- and lattice-distortion-dependent metamagnetic transitions have been demonstrated in layered systems [29, 31]. A spin-flop metamagnetic transition was demonstrated in $Sr_2IrO_4$ as a function of the S-L coupling energetics [29]. A stronger interlayer coupling is the theoretical basis of the spin-flop transition since the coupling strength changes from $Sr_2IrO_4$ to $Sr_3Ir_2O_7$ [31]. In the present CIO/CMO SLs, the resistance oscillations in the form of AMR arise from an oscillating magnetic moment embedded in a system with in-plane biaxial magnetic anisotropy. The sharp kink- and step-like transitions are superimposed on these oscillations, inducing an additional component of about 50% $\phi$-AMR, having the origin in spin-flop metamagnetic transition.

Based on the dimensionality, both CMO and CIO thin films exhibit a canted AFM phase with diminishing magnetic moments. In their SLs, the canted AFM transition at ~70–100 K suggests a modification of the electronic structure at the interface or the constituent layers. In SLs involving manganites, the valence state can be altered at the interfaces through the transfer of charge, producing emergent phenomena [36-39]. Charge transfer in $SrIrO_3/SrMnO_3$ SLs (large bandwidth analog of CIO/CMO SLs) yields interface FM phases. In order to explore the microscopic origin of the $\phi$-AMR in our CIO/CMO SLs, the interfacial charge transfer and electronic structure close to the Fermi level were qualitatively visualized via the peak energy and peak structure of the x-ray absorption spectroscopy (XAS) at both the Mn and Ir L-edges [40-42]. Samples chosen for this study comprise SLs with 2 UCs of one constituent and more than 4 UCs of the other constituent, along with $(MI44)_5$ having the same period in both layers. The spectra contain two features (Figures 4a, b), the $L_3(2p_{3/2})$ and $L_2(2p_{1/2})$ edges due to the spin-orbit coupling in the Mn 2p core hole. As evident in Fig. 4a, the total electron yield (TEY) mode spectra of all SLs have the same $L_3$ energy peak position for CMO layers with a period of 4 or more UCs. However, a prominent peak shift toward lower energy with respect to the other samples is evident for (MI25)$_5$. As the TEY detection mode is surface-sensitive (probes 3–5 nm thickness), we turn to the bulk-sensitive total fluorescence yield mode Mn-edge spectra (Fig. 4b). These show an energy shift of the L$_3$ edge similar to that of the TEY spectra. The shifting of the L$_3$ edge peak toward lower energy suggests the presence of Mn$^{3+}$ ions [6, 40-42], presumably formed by the transfer of charge from the Ir to the Mn at the interface.

The XAS at the Ir L-edge spectra (Figures 4c, d) depicts a shifting of the L-edge peak toward higher energy values with respect to an IrO$_2$ reference, which has an oxidation state of Ir$^{+4}$. This implies that the average Ir valency is more than +4, as expected for CIO due to the charge transfer with Mn [43-44]. Combined with the Mn edge spectra, it is clear that charge transfer from Mn$^{+4}$ to Ir$^{+4}$ takes place in all samples, while there also appears to be more charge transfer in thinner CMO layers. A nearly constant shift in Ir edge for all SLs suggests that CIO tends to lose only a constant fraction of its charge, which is absorbed by the CMO layer. In this case, the SLs with thinner CMO layers will receive a large fraction of electrons compared to their counterparts with thicker layers. This mechanism effects a larger shift in the Mn edge for (MI25)$_5$ with 2 UCs compared to other SLs in which CMO is thicker compared to CIO. Also, the larger thickness/volume of CIO (as in (MI25)$_5$) will transfer a larger number of electrons to CMO. Overall, the XAS data shown here support the crucial role of the thickness/volume of both layers in reconstructing the interfacial electronic structure of Mn. This electronic reconfiguration at the interface is consistent with the emergence of an anomalous magnetism in these SLs, which is more pronounced in the SLs with 2 UCs of CMO.

**Discussion**

The charge transfer evidently depends on the number of available carriers (or layers) of CIO and the number of CMO layers. The CMO with a severely distorted lattice has a strong affinity for electrons due to its vacant $e_g$ orbital near the Fermi level. In a different study on Ce$^{4+}$-doped CaMnO$_3$, even a 2–4% electron doping at the Mn-sites induced canting of the AFM lattice by 8°, increasing the magnetic moments by more than one order of magnitude [45]. In other relevant studies on CaRuO$_3$/CaMnO$_3$ SLs, the leakage of electrons into CaMnO$_3$ decays exponentially from the interface to the inner part of the layer [46-49]. Consequently, a fraction of Mn$^{3+}$ proportional to the electrons received at the interface and within the CMO layer induces a double-exchange-governed FM phase or a largely canted AFM phase at the CMO interface layer. The innermost

CMO layer will tend to remain AFM. Both theory and experiments agree on the formation of such magnetic gradient across the CMO layer. This was also explored in the present CIO/CMO by looking for exchange-bias fields ($H_{EB}$) that develop at the virtual FM and AFM phases of interface and bulk of CMO layers, respectively [47-50]. As seen in Fig. 4e, (MI22)$_{10}$, (MI33)$_5$, (MI44)$_5$, and (MI84)$_5$ exhibit a $H_{EB}$ of 3, 15, 50, and 35 Oe, respectively. As depicted in Fig. 4f, the first SL has its CMO layers only at the interface; thus, it lacks a virtual FM/AFM interface arrangement necessary for the $H_{EB}$. For the latter two SLs with the thicker CMO layers, the formation of a FM interface and AFM bulk manifests as $H_{EB}$. The amplitude of this $H_{EB}$ scales with the increase in CMO layer thickness.

Finally, we emphasize that the choice of the 3d compound is the key to yield the effective interlayer coupling and distortion required to tune a large $\phi$-AMR. In the case of the CIO/STO SLs, the $(a^-,a^-,c^+)$ octahedral distortions in the CIO layers propagate into the mediating STO layer, provided the STO layer consists of only one unit cell [8]. In contrast, in the present CIO/CMO SLs, this distortion propagates into the mediating CMO layer as thick as eight UCs. This suggests the CMO to be the most suitable 3d candidate known so far to promote interlayer exchange coupling. This is attributed to the fact that in the dimensional limit of a few unit cells, both CMO and CIO films are orthorhombically distorted with the same octahedral rotation pattern $(a^-,a^-,c^+)$ and have a similar in-plane Dzyaloshinskii–Moriya-type canted AFM phase [8, 51]. This structural similarity between the two constituents enhances the CIO interlayer coupling and results in a large biaxial anisotropy. The interlayer coupling in these SLs is such long-range that it persists even when the CIO layers are separated by 8 UCs of CMO. In addition, increasing the CMO layer thickness from 2 UCs to 8 UCs tunes the AMR by more than three orders of magnitude, underlining the adverse role of the anisotropic interfaces. We believe that the presence of strong $H_{EB}$ in thick SLs (MI44)$_5$ and (MI84)$_5$ creates graded AFM/FM phases within the CMO layers which impede a smooth CIO interlayer coupling. The role of $H_{EB}$ in hindering the interlayer coupling of CIO layers is also evident from the fact that an AMR of 25% at 20 K in (MI22)$_{10}$ (i.e. above spin-flop transition), albeit absent $H_{EB}$, reduces to just 3% in (MI33)$_5$ at 10 K, along with the moderate $H_{EB}$.

The CIO/CMO SLs we studied are proved to be the most efficient 3d-5d heterostructures for achieving an unprecedented AMR of ~70%, utilizing two key factors: a strong biaxial anisotropy and a spin-flop metamagnetic transition. The magnetic-anisotropy-driven AMR signals observed in this study are the strongest not only in the 3d-5d class of heterostructures but across all known complex oxide devices. A combined control of tilt pattern and SL construction is demonstrated to optimize the bi-axial magnetic anisotropy, the interlayer coupling mediated by a thick layer, and the field-pseudospin coupling. All these facets coalesce constructively to maximize the transport anisotropies in 3d-5d SLs. This proof-of-concept study is set to introduce new avenues for designing highly sensitive AMR readout devices for emerging AFM spintronics.

**Methods**

*Sample fabrication and structural characterization.* CMO/CIO films and superlattices were fabricated using pulsed laser deposition onto SrTiO$_3$ (STO) *(100)* substrates. The superlattices were deposited at a substrate temperature of 730°C and an oxygen partial pressure of 6 Pa followed by post-annealing at the same temperature and pressure for 30 minutes. The thickness of the superlattices were precisely controlled/monitored by *in-situ* reflection high-energy electron diffraction (RHEED). The high structural quality of the superlattices was examined in detail by performing room-temperature X-ray diffraction (XRD) $\theta$-$2\theta$ scans using a PANaylitcal X'pert Pro diffractometer. Out-of-the plane lattice parameters and strain states were studied by performing reciprocal space map (RSM) measurements, while the thicknesses of the superlattices were confirmed by X-ray reflectivity (XRR).

*Electrical characterization.* Transport and magnetotransport measurements were done in a Quantum Design Physical Properties Measurement System (PPMS) equipped with rotator module. The resistivity measurements were performed in constant-current mode at 10 µA to avoid Joule heating. The electrical contacts were made with Al wire using a wire bonder (West Bond make) in the four-probe geometry for the resistivity measurements. To obtain the angular dependence of the resistivity, the sample orientation was varied with respect to the magnetic field at a fixed temperature and magnetic field. For $\phi$-rotation, the magnetic field was applied parallel to (rotated within) the plane of the sample with the current applied along the *(hkl)* direction. For the $\theta$- and $\gamma$-rotations, the magnetic field was applied perpendicular to the sample surface, and the current was applied along the *(hkl)* directions.

*Magnetic characterization.* The magnetic behavior of the superlattices was investigated in a Quantum Design superconducting quantum interference device (MPMS-3). The temperature dependence of the magnetization was measured at various applied magnetic fields under the field-cooled protocol during the warming cycle. Magnetization isotherms were measured at various temperatures by sweeping the magnetic field between ± 7 T. In order to study the exchange bias effect in the superlattices, magnetization loops were obtained after cooling the samples in the presence of the applied magnetic field from the paramagnetic region to the temperature of the measurement.

***HAADF-STEM characterization.*** Cross-sectional high-resolution high-angle annular dark-field scanning transmission electron microscopy (HAADF-STEM) imaging was performed on a Talos F200X microscope (FEI) operated at an accelerating voltage of 200 kV. Prior to STEM analysis, the specimen mounted in a high-visibility low-background holder was placed for 8 s into a Model 1020 Plasma Cleaner (Fischione) to remove possible contamination. Classical TEM cross-sections of the SL structures glued together in face-to-face geometry using G2 epoxy glue (Gatan) were prepared by sawing (Wire Saw WS 22, IBS GmbH), grinding (MetaServ 250, Bühler), polishing (Minimet 1000, Bühler), dimpling (Dimple Grinder 656, Gatan), and final Ar ion milling (Precision Ion Polishing System PIPS 691, Gatan).

***X-ray absorption spectroscopy measurements.*** Room-temperature x-ray absorption spectroscopy (XAS) at the Mn L2,3-edges in both the total electron yield (TEY) mode and the total fluorescence yield (TFY) mode were performed at beamline 29-ID of the Advanced Photon Source (APS), ANL, USA. Measurements at the Ir L2,3 edges were performed at beamline 4-ID-D of the APS, ANL, USA. Data were collected in partial fluorescence yield mode by monitoring the Ir $L\alpha$ and $L\beta$ emission lines using an energy sensitive Si-drift detector.

Acknowledgements

D.S.R. thanks the Department of Science and Technology (DST) Nanomission for financial support under research project No. SM/NM/NS-84/2016 and the Science and Engineering Research Board (SERB) Technology, New Delhi under the project no. EMR/2016/003598. M.V. acknowledges the DST, India for the INSPIRE faculty award (DST/INSPIRE/04/2017/003059). XAS measurements were performed at UC San Diego in a search for materials for new spin-torque oscillators conducted by Quantum Materials for Energy Efficient Neuromorphic Computing, an Energy Frontier Research Center funded by the U.S. Department of Energy (DOE), Office of Science, Basic Energy Sciences (BES) under Award # DE-SC0019273. This research used resources of the Advanced Photon Source (29ID for soft XAS and 4-ID-D for hard XAS), a U.S. Department of Energy (DOE) Office of Science User Facility operated for the DOE Office of Science by Argonne National Laboratory under Contract No. DE-AC02-06CH11357; with additional support by National Science Foundation under Grant no. DMR-0703406. Extraordinary facility operations were supported in part by the DOE Office of Science through the National Virtual Biotechnology Laboratory, a consortium of DOE national laboratories focused on the response to COVID-19, with funding provided by the Coronavirus CARES Act. Authors thank Mr. G. L. Prajapati and Mr. Manoj Prajapat for their help in performing magnetization, transport and X-ray diffraction characterizations as well as R. Aniol for TEM specimen preparation. Furthermore, the use of the HZDR Ion Beam Center TEM facilities and the funding of TEM Talos by the German Federal Ministry of Education of Research (BMBF; grant No. 03SF0451) in the framework of HEMCP are acknowledged.

Author contributions

D.S.R. conceived and supervised this project. S.S., M.V. and T.T. prepared samples, performed the structural, magnetic and magnetotransport measurements and analyzed data with inputs from D.S.R. P.P. and R.H. performed HAADF-STEM characterizations. XAS measurements were performed by S.D., B.G., A.F., F.R., G.F., Y.C., and D.H.. D.S.R. and M.V. drafted the manuscript with contributions from A.F., S.S., S.D. and R.H. All authors discussed the data and contributed to the manuscript.

Competing interests:
Authors declare no competing interests.

Figure 1:

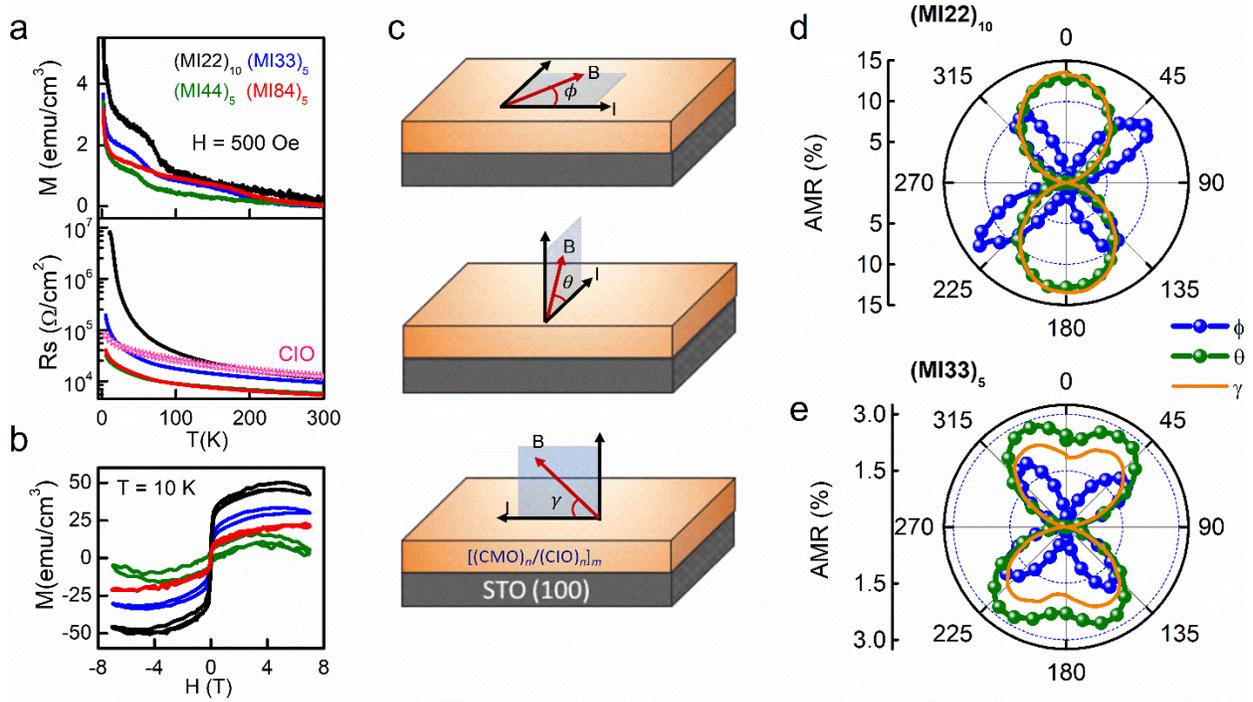

**Fig. 1 Comparison of $\phi$, $\theta$, and $\gamma$-AMRs in CMO/CIO superlattices.** **a** Temperature dependence of magnetization and sheet resistance for (MIxy)$_z$ (x = y = 2-4) along with (MI84)$_5$ to elucidate the effect of a larger CMO period. **b** M (H) plots for (MIxy)$_z$ (x = y = 2-4) and (MI84)$_5$ at T = 10 K. **c** Schematic illustration of three different rotational geometries to measure the AMR derived from the angle between the current (I) and the applied magnetic field (B) with respect to the plane of rotation. Polar plots presenting the comparison of $\phi$, $\theta$, and $\gamma$-AMR for **d** (MI22)$_{10}$ and **e** (MI33)$_5$ superlattices. AMR: anisotropic magnetoresistance; CMO: CaMnO$_3$; CIO: CaIrO$_3$. (MIxy)$_z$: M and I refer to CMO and CIO layers, respectively, x, y = number of unit cells/period, z = repetitions.

Figure 2:

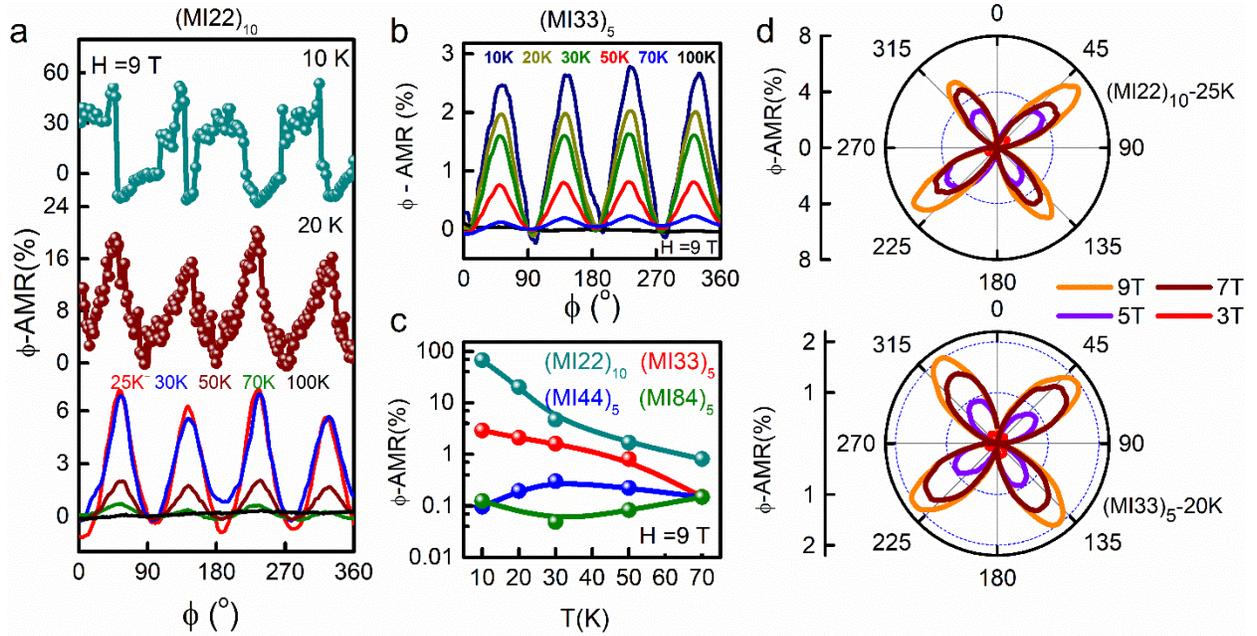

**Fig. 2 Temperature and magnetic field dependence of the $\phi$-AMR.** $\phi$-AMR for **a** $(MI22)_{10}$, **b** $(MI33)_5$ measured at various temperatures at H = 9 T. **c** Variation in the $\phi$-AMR amplitude as a function of temperature for H = 9 T for various superlattices. **d** Polar plots of the $\phi$-AMR highlighting its field dependence for the $(MI22)_{10}$ and $(MI33)_5$ superlattices at T = 25 K and 20 K, respectively. AMR: anisotropic magnetoresistance. $(MIxy)_z$: M and I refer to $CaMnO_3$ and $CaIrO_3$ layers, respectively, x, y = number of unit cells/period, z = repetitions.

Figure 3:

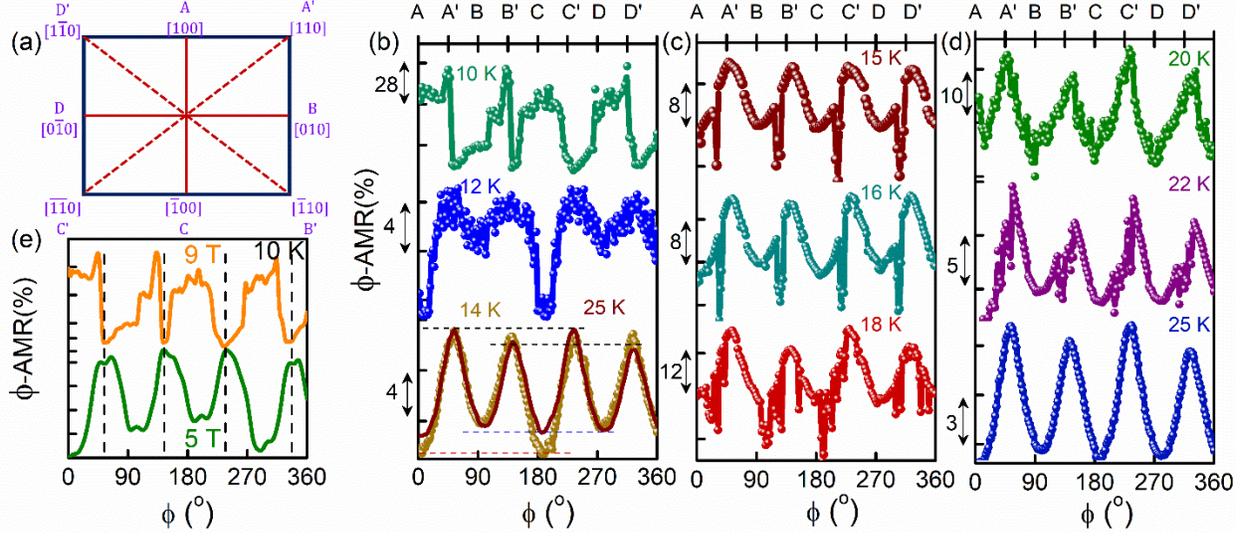

**Fig. 3 A spin-flop transition manifesting in the AMR of (MI22)$_{10}$. a** <100> crystallographic in-plane family of directions are labelled as A, B, C, and D and the <110> family as A′, B′, C′, and D′. The former corresponds to the low-resistance state, while the latter shows the peak resistance. **b–d** $\phi$-AMR of the (MI22)$_{10}$ SL in the temperature range of 25 K – 15 K at a magnetic field of 9 T, manifesting the onset of the spin-flop transition at 22 K evident from the additional four-fold symmetry in $\phi$-AMR. **e** Spin-flop-transition-induced reversal in the polarity of the peak amplitude in $\phi$-AMR with the increase in magnetic field for the (MI22)$_{10}$ SL at 10 K. AMR: anisotropic magnetoresistance. (MIxy)$_z$: M and I refer to CaMnO$_3$ and CaIrO$_3$ layers, respectively, x, y = number of unit cells/period, z = repetitions.

Figure 4:

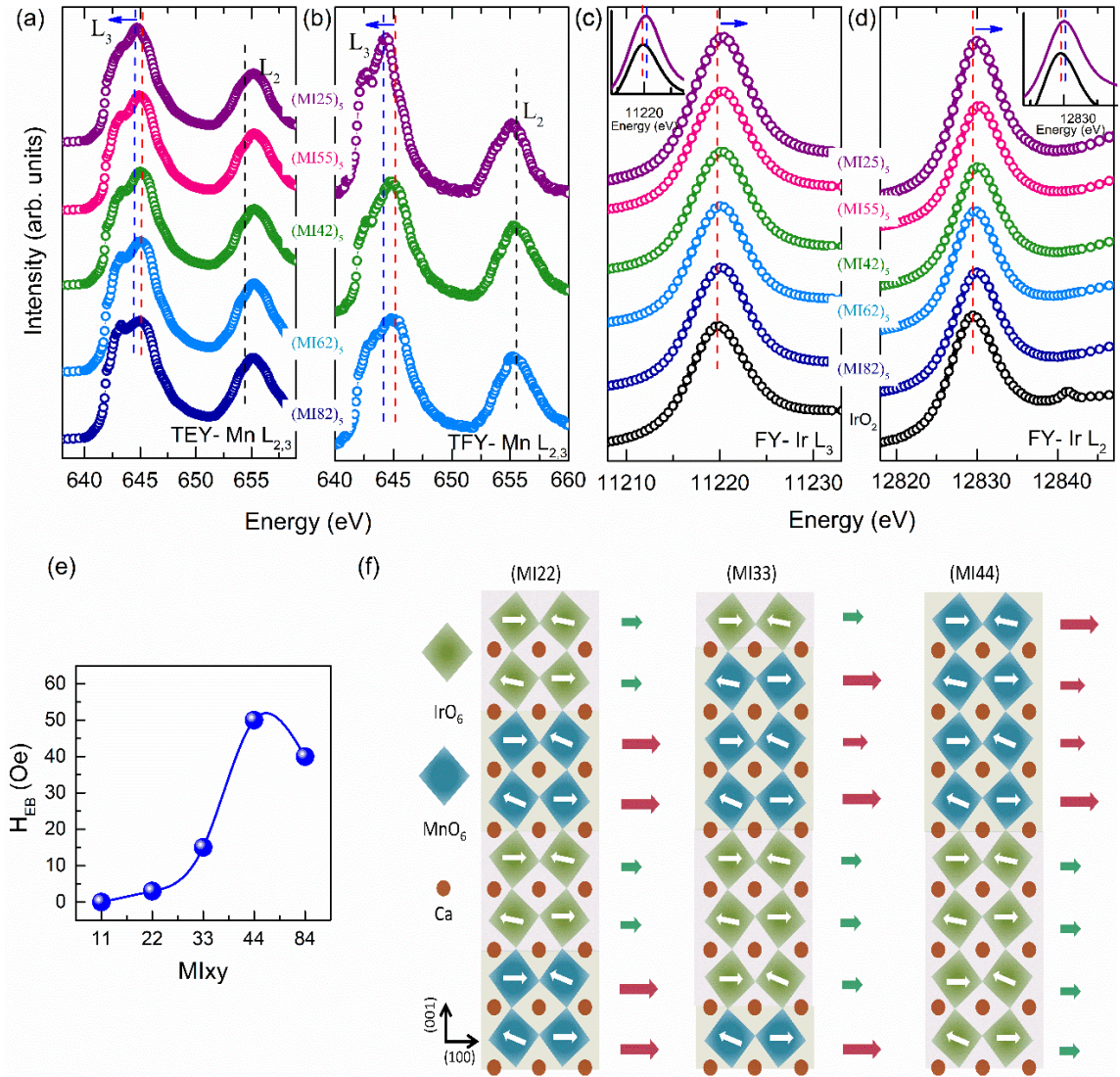

**Fig. 4 Element-specific charge transfer and mechanism of interfacial magnetization.** XAS spectra around the Mn $L_{2,3}$ edges for $(MIxy)_z$ measured in **a** TEY and **b** TFY mode. XAS spectra of the **c** Ir $L_3$-edge **d** Ir $L_2$-edge for $(MIxy)_z$ measured in the fluorescence yield modes. The insets depict the comparison of the $(MI25)_5$ sample with the $IrO_2$ reference **e** The strength of exchange bias field with the varying stacking of CMO and CIO layers in the (MIxy) superlattices. **f** Schematic illustrating the canted magnetic structure along the <001> direction for the superlattices MI22, MI33 and MI44. The rotation of adjacent $IrO_6$ and $MnO_6$ octahedra and the charger transfer at the interface give rise to a larger net ferromagnetic moment at the CMO/CIO interface. TEY: total electron yield; TFY: total fluorescence yield; XAS: x-ray absorption spectroscopy; AMR: anisotropic magnetoresistance; CMO: $CaMnO_3$; CIO: $CaIrO_3$. $(MIxy)_z$: M and I refer to CMO and CIO layers, respectively, x, y = number of unit cells/period, z = repetitions.

Supplementary Information

# Giant anisotropic magnetoresistance with dual-four-fold symmetry in CaMnO₃/CaIrO₃ heterostructures

I. **Sample details:**

The CaMnO$_3$(CMO)/CaIrO$_3$(CIO) SLs were coded as **(MIxy)$_z$**, where M and I refer to the CMO and CIO layers, respectively, and were distinguished in different categories based on the periods of the constituent layers, as:

A) *Simultaneous variation of CMO and CIO unit cells*:

(MIxy)$_z$ for x = y = 2-4, as (MI22)$_{10}$, (MI33)$_5$, and (MI44)$_5$, along with (MI84)$_5$ with larger CMO period

B) *CMO control of AMR by fixing the CIO period:*

*i)* (MIx2)$_5$ for x = 2, 4, 6, and 8 as (MI22)$_{10}$, (MI42)$_5$, (MI62)$_5$, and (MI82)$_5$, fixing the CIO period to 2 UCs;

*ii)* (MIx4)$_5$ for x = 2, 4, and 6 as (MI24)$_5$, (MI44)$_5$, and (MI64)$_5$, fixing the CIO period to 4 UCs;

C) *CIO control of AMR by fixing the CMO period:*

*iii)* (MI2y)$_z$ for y = 2, 4, and 5 as (MI22)$_{10}$, (MI24)$_5$, and (MI25)$_5$, fixing the CMO period to 2 UCs;

*v)* (MI5y)$_z$ for y = 2, 5, and 8 as (MI52)$_7$, (MI55)$_5$, and (MI58)$_5$, fixing the CMO period to 5 UCs.

## II. Structure:

All the SLs were prepared via pulsed-interval epitaxy using a RHEED-assisted pulsed laser deposition technique. Structure and sample quality were analyzed by various types of x-ray diffraction and HAADF-STEM imaging. The x-ray diffraction patterns as well as the x-ray reflectivity curves for the primary series of SLs $(MIxy)_z$ for $x = y = 2$-4 and $(MI84)_5$ are shown in figure S1. The reflectivity data for $(MI33)_5$ and other higher-period SLs suggest near atomically sharp interfaces. The quality was further confirmed by HAADF-STEM imaging of the thicker SL $(MI84)_5$. The data are shown of figure S2 on different length scales. While the overall thickness of the individual layers is in agreement with the nominal SL structure, the formation of step-like patterns suggests that there might be tiny thickness variations by one unit cell. It is noted that the formation of such steps is beyond a period of 20 unit cells, thus excluding a dominant effect. Further, it is also noted that the x-ray reflectivity of $(MI22)_{10}$ does exhibit the higher quality expected for a superlattice with 10 layers. This is elaborated in the next section.

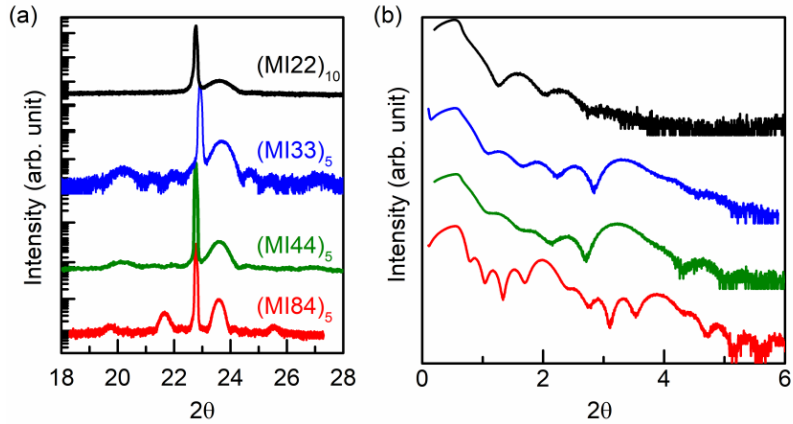

Figure S1: (a) $\theta$-$2\theta$ scans and (b) X-ray reflectivity for $(MIxy)_z$ with $x = y = 2$-4, i.e. $(MI22)_{10}$, $(MI33)_5$, and $(MI44)_5$, along with $(MI84)_5$.

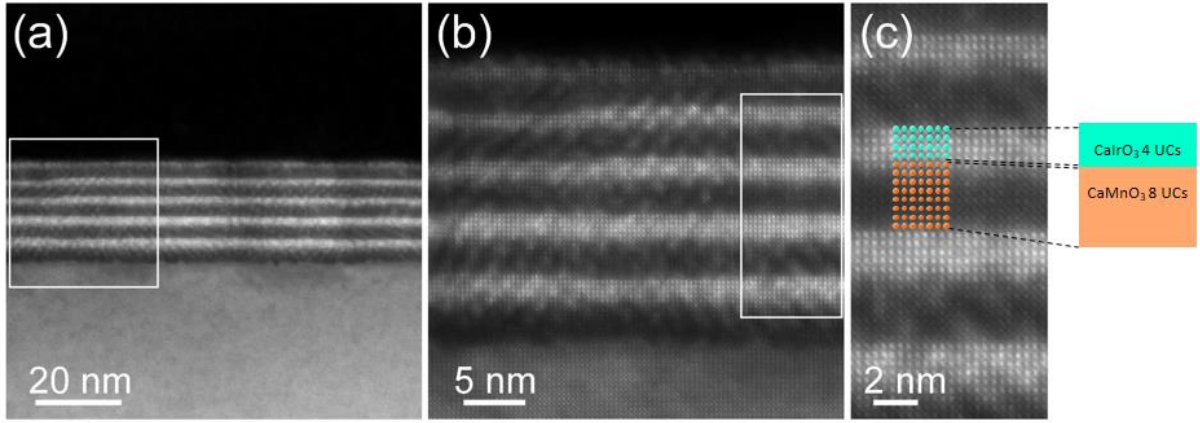

Figure S2: Cross-sectional high-resolution HAADF-STEM image showing the $CaMnO_3$ and $CaIrO_3$ layers in the $(MI84)_5$ superlattice. The white square/rectangle in panel (a)/(b) marks the image region displayed in panel (b)/(c), respectively.

III. **Choice of the $(MI22)_{10}$ SL based on $\phi$-AMR:**

In our attempts to achieve high-quality (MI22) SLs, several samples with a wide range of quality were prepared. In figure S3, the x-ray reflectivity of four such SLs suggests a clear difference in their quality. The number and quality of fringes of the first three with the same periodicity labeled as, $(MI22)_{10}$-A, $(MI22)_{10}$-B, and $(MI22)_{10}$-C, suggests that these SLs are of very high, moderate and poor quality, respectively. Their thickness as measured from the same data was found to be in the range of 14.5 – 17 nm, which is very close to the expected value of 15.6 nm. The fourth sample with lower periodicity $(MI22)_8$ is also of a good quality. Figure S3 also shows their maximum in-plane $\phi$-AMR amplitudes with a four-fold symmetry and well-defined sinusoidal behavior. While $(MI22)_{10}$-A and $(MI22)_{10}$-B exhibit a similar amplitude of ~ 8-9%, the same for the poor-quality $(MI22)_{10}$-C SL drops to ~ 2-3%. The $(MI22)_8$, a good-quality SL, exhibits an AMR of ~6% but with a lower signal-to-noise ratio. Given that both $(MI22)_{10}$-A and $(MI22)_{10}$-B exhibit the same amplitude of a sinusoidal four-fold $\phi$-AMR, we chose the latter for the main study due to a very emphatic reason. This SL also exhibits another sharp step-like four-fold component of $\phi$-AMR (of amplitude ~50%), which is superimposed on the sinusoidal $\phi$-AMR, thus revealing a unique *dual-four-fold* symmetry component of $\phi$-AMR in any class of SLs. The step-like four-fold AMR component is attributed to the spin-flop transition. The reason of spin-flop-induced giant $\phi$-AMR in $(MI22)_{10}$-B is understood as an interface-defect-induced reduction in the magnetic-field required to trigger this transition. A large number of such defects, as appears to be the case for the poor-quality (MI22)$_{10}$-C SL, become counterproductive, as its sinusoidal $\phi$-AMR falls significantly. Hence, it is the balance of acceptable quality and amplitude of $\phi$-AMR that (MI22)$_{10}$-B was chosen over (MI22)$_{10}$-A. Furthermore, the (MI22)$_8$ seems to behave like (MI22)$_{10}$-B. Below a temperature of 10 K for the former and 20 K for the latter sample, a sudden rise in $\phi$-AMR with reduced signal-to-noise ratio marks the onset of spin-flop-induced AMR. In (MI22)$_8$, however, detailed features on this transition could not be collected because of its large resistance below 8 K.

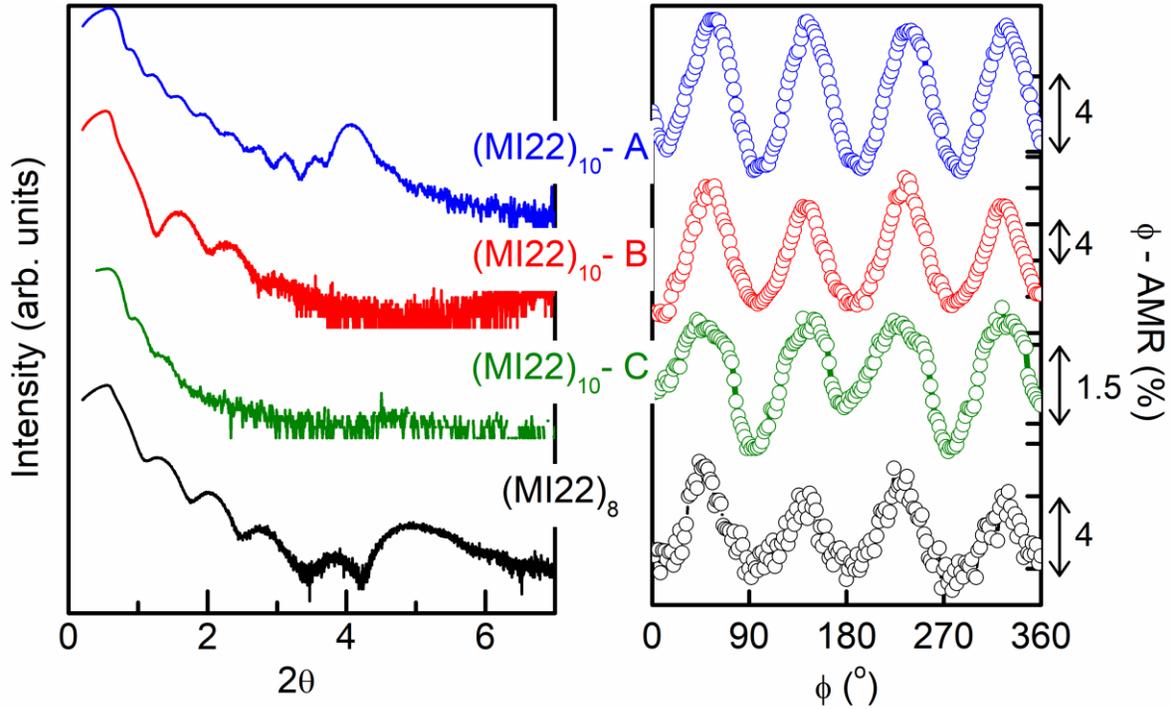

Figure S3: X-ray reflectivity curves and maximum $\phi$-AMR for (MI22)$_8$ and (MI22)$_{10}$ with different quality.

## IV. $B^2$-field dependence of the $\phi$-, $\theta$-, and $\gamma$ – AMR amplitude for (MI22)$_{10}$ and (MI33)$_5$:

The $\phi$–AMRs does not vary linearly with $B^2$ for any of CIO/CMO SL. Figure S4 depicts this behavior for two key samples, (MI22)$_{10}$ and (MI33)$_5$. This rules out the Lorentz scattering as underlying mechanism for this property.

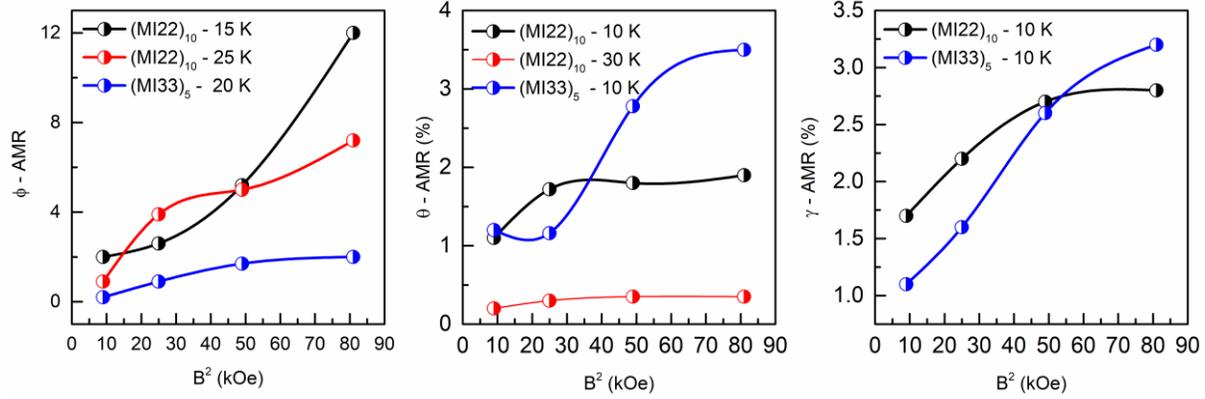

Figure S4: $\phi$-, $\theta$- and $\gamma$ – AMRs exhibiting non-quadratic dependence on B for the $(MI22)_{10}$ and $(MI33)_5$ SLs at various temperatures.

## V.     CMO influence on AMR by fixing the CIO period:

*i)* $(MIx2)_5$ for x = 4, 6, and 8 as $(MI42)_5$, $(MI62)_5$, and $(MI82)_5$, with fixing the CIO period to 2 UCs: The $\phi$-AMR values of this series of SLs is plotted in figure S5 below. It is clear that except for x=2, $\phi$-AMR of all other SLs is in range of 0.4-0.8%, suggesting that a short CMO period results in a large AMR value only if the CIO period is small too.

*ii)* $(MIx4)_5$ for x = 2, 4, and 8 as $(MI24)_5$, $(MI44)_5$, and $(MI84)_5$, with fixing the CIO period to 4 UCs: In this case, the variation of the CMO period for a fixed larger CIO period (4 UCs) shows a $\phi$-AMR in the range of ~ 0.02 – 1.5%; however, except for $(MI24)_5$, the AMR of the SLs is lesser than 0.2% These data are plotted in figure S5.

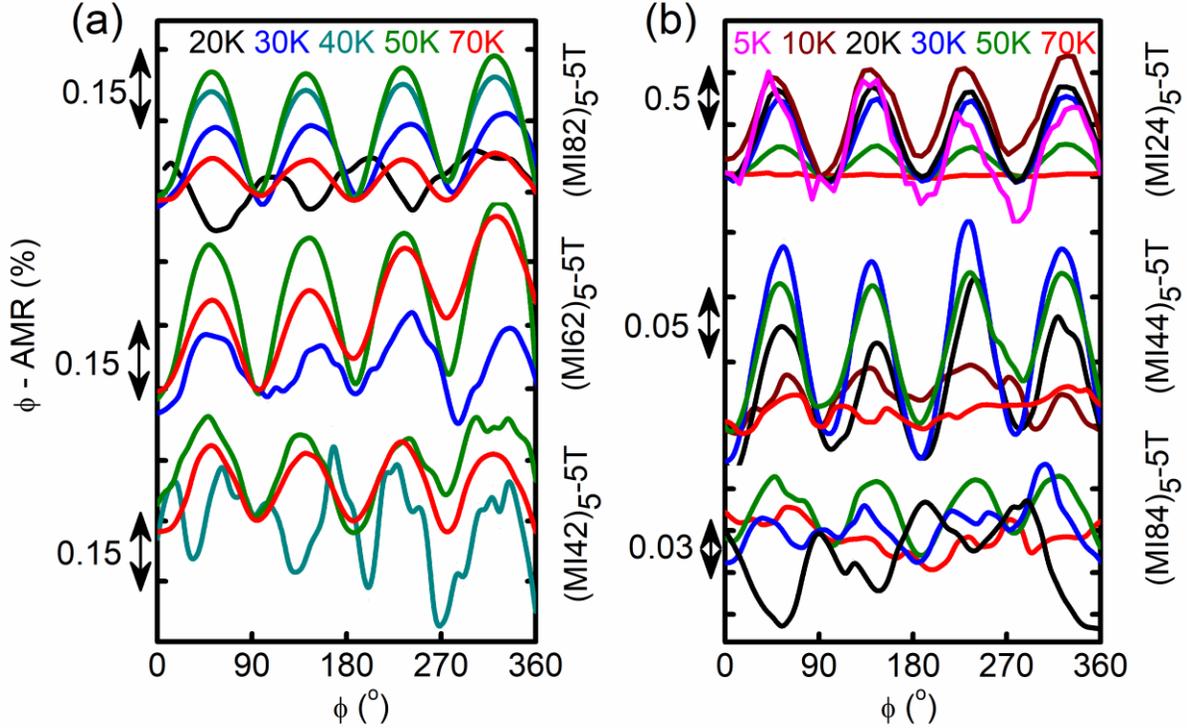

Figure S5: $\phi$-AMR for (a) (MIx2)$_5$ for x = 4, 6, and 8 (b) (MIx4)$_5$ for x = 2, 4, and 8 measured at various temperatures at H = 5 T.

## VI. CIO influence on AMR by fixing the CMO period:

*i)* (MI2y)$_z$ for y = 4, and 5 as (MI24)$_5$, and (MI25)$_5$, fixing the CMO period to 2 UCs: The $\phi$-AMR data for these SLs is plotted in figure S6(a). The AMR decreases significantly as the CIO period increases from 2 to 4 and 5.

*ii)* (MI5y)$_z$ for y = 2, 5, and 8 as (MI52)$_7$, (MI55)$_5$, and (MI58)$_4$, fixing the CMO period to 5 UCs: Here, for a thicker fixed CMO period, the decrease in AMR is so pronounced that for (MI58)$_5$, $\phi$-AMR follows the behavior of a CIO film of equivalent thickness. These data are plotted in figure S6(b).

iii) Similarity of AMR for (MI58)$_5$ and CIO: The AMR of (MI58)$_4$ having a large CIO period (of 8 UCs) is found to match with that of a 10-nm-thick CIO film, both in terms of phase and amplitude. This suggests that beyond a certain thickness of the CIO layers in the SLs (8 UCs in this case), the SL loses the biaxial anisotropy required for interlayer coupling, hence, it behaves like a single CIO film. A comparison of the AMRs of the two cases is plotted in figure S7.

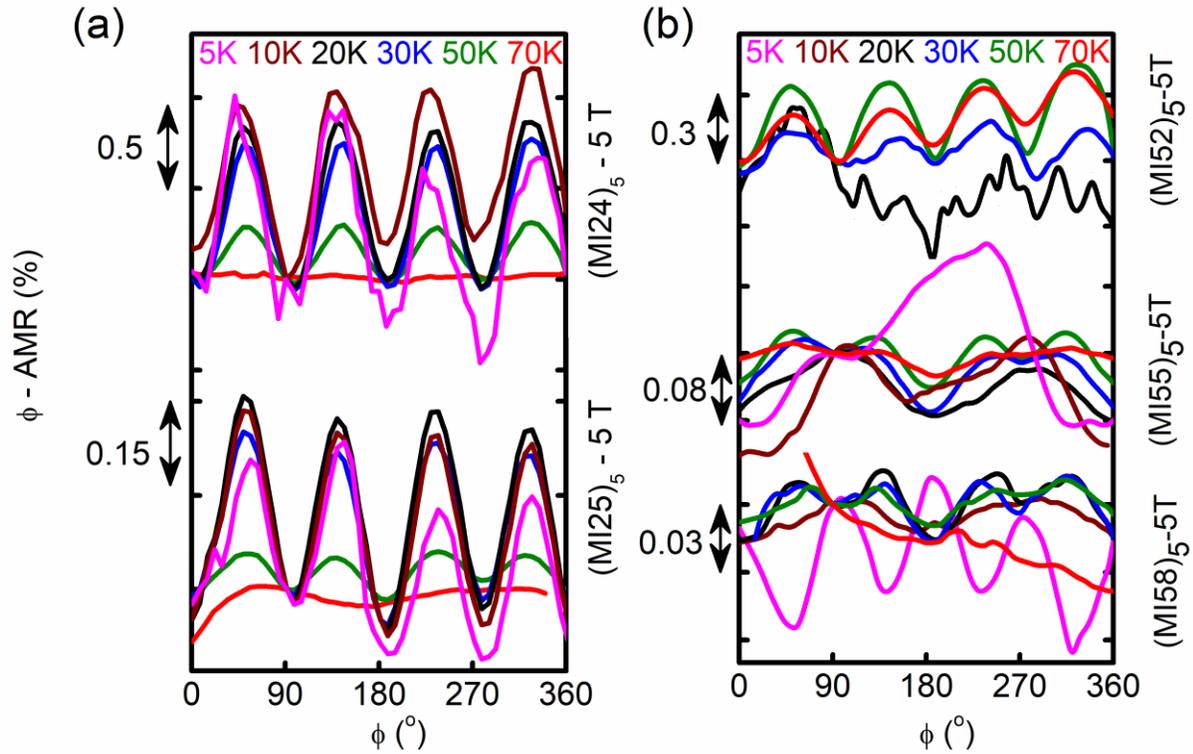

Figure S6: $\phi$-AMR for (a) (MI2y)$_z$ for y = 4 and 5 (b) (MI5y)$_z$ for y = 2, 5, and 8 measured at various temperatures at H = 5 T.

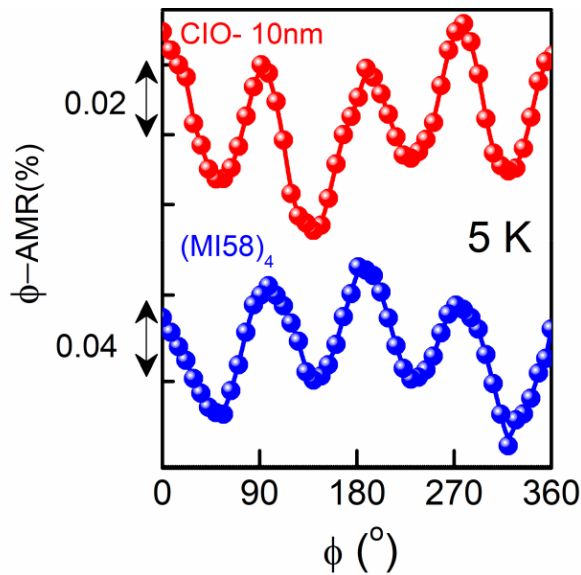

Figure S7: Comparison of $\phi$-AMR for a 10-nm-thick CIO film and the (MI58)$_4$ SL measured at T = 5 K.

## VII. Temperature dependence of $\phi$-AMR in context of magneto-elastic coupling:

$\phi$-AMR as a function of temperature for several SLs is plotted in figure S8 to find a correlation of the AMR amplitude with the magneto-elastic coupling. It is seen that for the samples exhibiting a large AMR, say for $(MI22)_{10}$, $(MI33)_5$, and $(MI24)_5$, the AMR amplitude peaks at the lowest temperatures (as measured around 5-10 K). However, as the AMR drops for other SLs, the peak shifts to higher temperatures in the range of 30-50 K. In iridates, it is known that the magneto-elastic coupling is maximum in the vicinity of the magnetic transition, which allows an easy rotation of the orthorhombic distortion with the magnetic field.

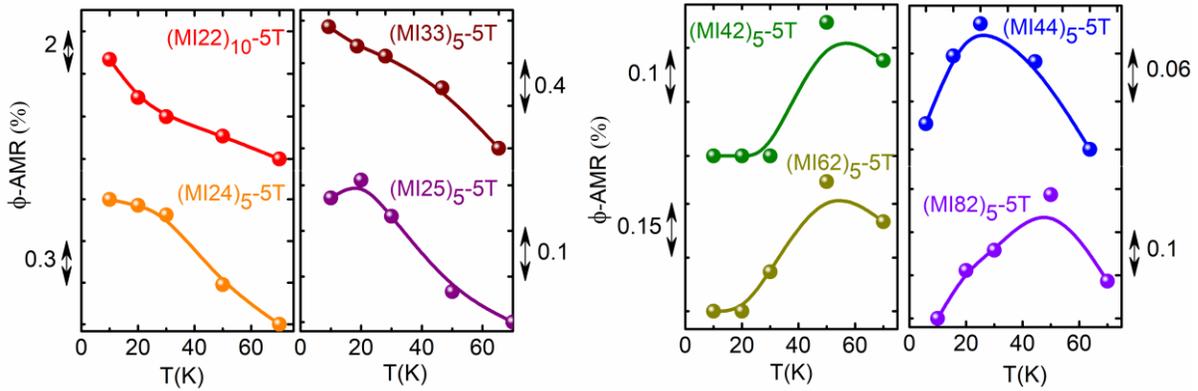

Figure S8: Temperature dependence of the $\phi$-AMR amplitude for $(MIxy)_z$ superlattices at H = 5 T.

## VIII. Magnetic-field dependence of $\phi$-AMR in $(MI22)_{10}$

A magnetic-field dependence of $\phi$-AMR at 15 K and 25 K was performed to elucidate the information about spin-flop-induced transition. It is seen in Figure S9 that the kink-like transition, as identified by spin-flop transition, occurs only at 9 T at 15 K. As expected, this feature is absent at 25 K.

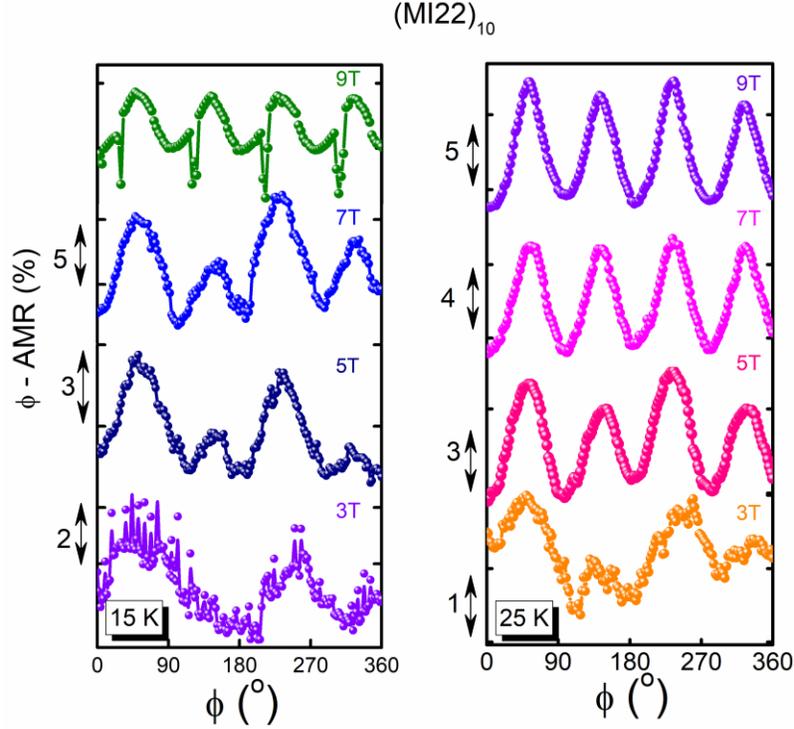

Figure S9: Field dependence of $\phi$-AMR for $(MI22)_{10}$ measured at T = 15 K (left) and T = 25 K (right).

IX. **Evidence of transition to a step-like metamagnetic transition evaluated by $\phi$-AMR in $(MI22)_{10}$:**

As seen in figure S10, a smooth sinusoidal $\phi$-AMR of around 10% at 14 K transforms to a giant 70% step-like spin-flop metamagnetic transition at 10 K. The spin-flop nature of this transition is evident from the fact that the trough of $\phi$-AMR at 14 K matches with the crest at 10 K. Matching the $\phi$-AMR at these two temperatures, it is clear that there is a sudden step-like drop in resistance at 10 K at an angle where the peak of the sinusoidal $\phi$-AMR starts setting in at 14 K. Correlating this $\phi$-AMR to the underlying magnetism, it is clear that spin-flop metamagnetic transition is responsible for the giant $\phi$-AMR at 10 K.

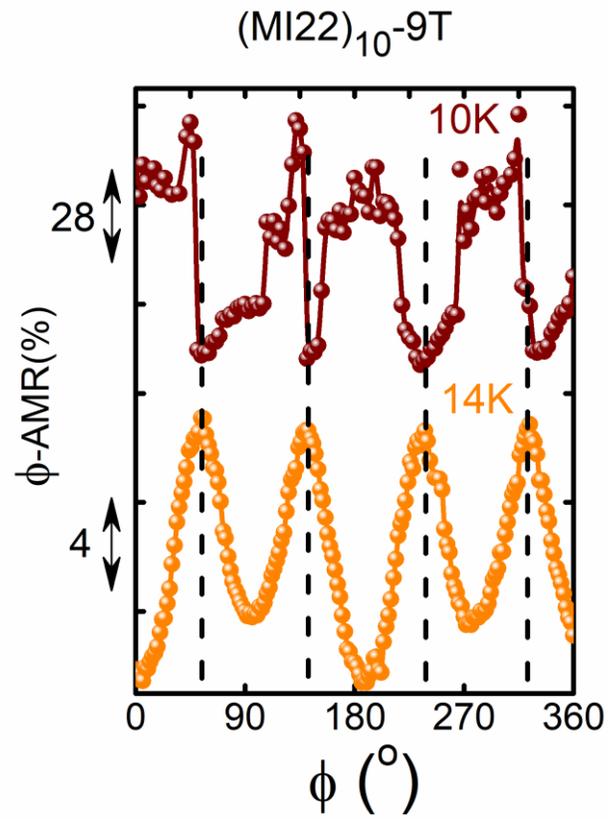

Figure S10: Comparison of $\phi$-AMR for $(MI22)_{10}$ at T = 10 K and T = 14 K measured at H = 9 T.